\documentclass[twocolumn,aps,showpacs]{revtex4}
\usepackage{graphicx}
\usepackage{amsmath}
\usepackage{amsfonts}
\usepackage{amssymb}

\begin{document}

\title{\textbf{The Branched Polymer Growth Model Revisited}}
\author{Ubiraci P. C. Neves $^{1}$}
\email{ubiraci@ffclrp.usp.br}
\author{Andr\'e L. Botelho $^{1}$}
\author{Roberto N. Onody $^{2}$}
\affiliation{$^{1}$Departamento de F\'{\i}sica e Matem\'{a}tica,
Faculdade de Filosofia, Ci\^{e}ncias e Letras de Ribeir\~{a}o Preto \\
Universidade de S\~{a}o Paulo - Ribeir\~{a}o Preto - SP, Brazil}
\affiliation{$^{2}$Departamento de F\'{\i}sica e Inform\'atica,
Instituto de F\'{\i}sica de S\~ao Carlos, \\
Universidade de S\~ao Paulo - S\~ao Carlos-SP, Brazil}

\begin{abstract}
The branched polymer growth model (BPGM) has been employed to
study the kinetic growth of ramified polymers in the presence of
impurities. In this article, the BPGM is revisited on the square
lattice and a subtle modification in its dynamics is proposed in
order to \textit{adapt} it to a scenario closer to reality and
experimentation. This new version of the model is denominated the
\textit{adapted branched polymer growth model} (ABPGM). It is
shown that the ABPGM preserves the functionalities of the monomers
and so recovers the branching probability $b$ as an input
parameter which effectively controls the relative incidence of
bifurcations. The critical locus separating infinite from finite
growth regimes of the ABPGM is obtained in the $(b,c)$ space
(where $c$ is the impurity concentration). Unlike the original
model, the phase diagram of the ABPGM exhibits a peculiar
reentrance.


\end{abstract}

\maketitle

\section{Introduction}
The statistical mechanics of polymers has become a very prolific
research area. During the last decades, many models have been
proposed to describe the configurational properties of linear and
branched polymers and a variety of methods have been employed to
study these systems. The classic model for a linear polymer in
dilute solution (or at high temperature) is the self-avoiding walk
(SAW) \cite{flory53,gennes79}. On the other hand, connected
clusters (lattice animals) provided a good model for branched
polymers in the dilute limit \cite{lubensky79}.

The kinetic growth walk (KGW) was proposed as an alternative model
to describe the irreversible growth of linear polymers [4-6]. Like
the SAW, every KGW chain does not intercept itself. But whereas in
a SAW the next step is randomly chosen from among \textit{all}
nearest-neighbor sites (excluding the previous one), in a KGW the
choice is among the \textit{unvisited} sites \cite{majid84}.
Therefore the KGW is less sensitive to attrition. Besides,
although both models span the same set of configurations, the
N-step SAW chains are \textit{all} equally weighted while the
statistical weights of the KGW chains can be different.
Nevertheless, it was shown that both models present the same
critical exponents \cite{lyklema86}.

Later, Lucena \textit{et al.} \cite{lucena94} generalized the KGW
in order to allow for branching as well as for impurities. This
generalized model (which became known as the branched polymer
growth model - BPGM) was found to exhibit an interesting phase
transition (due to competition between hindrances and branching)
separating infinite from finite growth regimes \cite{lucena94}. In
the following years, several authors have studied the BPGM [8-15].
The topological and dynamical aspects of the model were
investigated \cite{neves95}. Besides, the system was shown to
achieve the criticality through a self-organization growth
mechanism \cite{andrade97} and to exhibit a transition from rough
to faceted boundaries for large values of the branching
probability \cite{neves95,lucena99}. The model was also studied
through an exact enumeration of bond trees and ergodicity
violation was discussed \cite{onody00}. The question of the
universality class of the BPGM was also investigated. In contrast
to the common belief that branched polymers belong to the
universality class of lattice animals
\cite{lubensky78,havlin84}, the study of the growth process in chemical $l-$%
space with estimates of structural exponents led to the proposal
that the BPGM belongs to the universality class of percolation
\cite{bunde95,porto96}. On the Bethe lattice, this proposal is
based on analytical results \cite{porto96}. A further analysis
using finite-size scaling techniques led to the conclusion that
the BPGM is \textit{not} in the same universality class of
percolation in two dimensions \cite{botelho00}.

In the present paper, we revisit the BPGM on the square lattice.
We point out that, although the input parameter $b$ is named
branching probability, it does not have effective control of the
relative incidence of bifurcations (branches) in the simulated
polymers. This seems to be an undesirable aspect of the model
since the degree of branching is an important quantity that is
usually measured in real ramified polymers. Besides, according to
the growth
rules of the BPGM, a monomer which is chosen to bifurcate (with probability $%
b$) but has only one empty nearest neighbor site is compelled to
grow linearly thus changing its functionality from $3$ to $2$. In
real experiments this change is not allowed since the
concentrations of polymers with different functionalities are
fixed. In order to preserve the functionalities of the monomers as
well as to recover the meaning of the branching probability $b$,
we propose a subtle modification in the dynamics of the BPGM. So
this new version of the model aims to \textit{adapt} it to a more
realistic scenario and is called the \textit{adapted branched
polymer growth model} (ABPGM). In the following sections, we
define the ABPGM and present the phase diagram in the $(b,c)$
space. This diagram is found to exhibit a peculiar reentrance. We
introduce the concept of \textit{frustration} and compare the
\textit{effective branching rate }$b_{ef}$ with the input
parameter $b$ for both the original and adapted models. Our
results concerning the ABPGM are shown to be much closer to the
ideal behavior $b_{ef}=b$. Finally, we present a discussion based
on the clusters topologies of both models.

\section{The Adapted Branched Polymer Growth Model}
First of all, let us review the BPGM. Consider a $L\times L$
square lattice with a certain concentration  $c$ of sites randomly
filled by impurities. At the initial time $t=0$, the polymer
starts growing from a monomer seed located at the center of the
lattice towards a random empty nearest-neighbor site. At time
$t=1$, this chosen site is occupied by a monomer (growing tip)
which now \emph{may} bifurcate or follow in one direction (linear
growth). The growth directions are randomly chosen among the
available ones (those which lead to empty neighbors) in a
clockwise way. At every time $t$ ( $t\geq 2$), the process is
repeated for all actual monomeric ends following the sequence of
their appearances. If a polymer end has \emph{no} empty
nearest-neighbor site then it is trapped in a \textquotedblleft
cul de sac\textquotedblright\ and stops growing (it is then called
a \emph{dead end}). If only one adjacent site is available then
the linear growth is obligatory. If at least two adjacent sites
are empty then the growing end bifurcates with probability $b$ or
follows linearly with probability $1-b$. In a particular
experiment of the BPGM, the polymer growth is simulated according
to the above rules. Depending on the values of parameters $b$ and
$c$, the polymer can either grow indefinitely or stop growing
(\emph{die}) at a finite time if all its current tips are dead
ends. The experiment is finished either when the polymer touches
the frontier of the lattice (\emph{infinite} polymer) or when it
dies (\emph{finite} polymer). The ensemble over which averages are
performed is constituted by a great number $N_{e}$ of experiments.

Each polymer configuration of the BPGM can be identified with a
self-avoiding loopless graph (\emph{bond tree}) although the
reciprocal is not always true \cite{neves95}. Thus the BPGM can be
mapped into a particular \emph{subset} of the ensemble of all
possible bond trees. These tree graphs have been applied in models
of branched polymers and can be classified by the number $N$ of
bonds and the number $N_{k}$ of vertices with $k$ bonds
\cite{camacho92} . In a typical finite polymer configuration of
the BPGM, one may find twofold and threefold sites representing
monomers of functionalities $2$ and $3$, respectively. The monomer
seed and dead ends represent monofunctional monomers.
Tetrafunctional units are impossible in this model once
trifurcations are not allowed. For a tree graph generated by the
BPGM one has the following constraints \cite{camacho92}:

\begin{eqnarray}
\sum_{k=1}^{3}N_{k}=N+1
\end{eqnarray}

and

\begin{eqnarray}
    \sum_{k=1}^{3}kN_{k}=2N \ .
\end{eqnarray}

The occurrence of bifurcations in the model determines the
relative incidence of trifunctional monomers (branches). This is
basically the \emph{degree of branching} of the polymer, an
important quantity that is
usually measured in real ramified polymers. However, although the parameter $%
b$ is named \emph{branching probability}, we shall see that it
does \emph{not} have effective control of the relative amount of
branches (bifurcations) in the BPGM. Indeed, according to the
rules of the BPGM, the parameter $b$ is not the probability that
any tip bifurcates but instead it is the conditional probability
that a tip \emph{with two or more empty nearest neighbors}
bifurcates. Every tip with just one vacant nearest-neighbor site
grows linearly with probability equal to one. From another
viewpoint, it would also be desirable to consider that any monomer
that is linking to the polymer has an effective probability $b$ of
being a trifunctional unit. Even if this were considered in the
BPGM, a problem would become evident. Any growing end chosen (with
probability $b$) as a trifunctional monomer would \emph{not} be
able to bifurcate if only one vacant site were available. In this
case, the functionality of the monomer would not be respected
since (according to the BPGM) the growing end would forcibly
follow a linear growth like a bifunctional monomer.

In order to preserve the functionalities of the monomers we
propose a modification in the dynamics of the BPGM. Besides to
make the model a little more realistic, our proposal turns the
branching probability $b$ into an effective control parameter of
the relative frequency of bifurcations. So it \emph{adapts} the
model to a scenario closer to reality and experimentation. We
denominate it the \emph{adapted branched polymer growth model}
(ABPGM).

The ABPGM is defined just like the BPGM except for the following
differences:

\begin{enumerate}
\item In the process of polymerization, let a \emph{free end} (a
growing tip with at least one empty nearest-neighbor site) be a
trifunctional monomer with probability $b$ or a bifunctional
monomer with probability $1-b$. If the free end is a trifunctional
monomer but there is only one empty nearest neighbor site then it
stops growing and becomes a dead end.

\item At every time unit, all current sites on the front of growth
of the polymer are visited in a \emph{random} sequence.
\end{enumerate}

We remember that in the BPGM, a free end with \textit{just one}
empty nearest neighbor grows linear with probability one while in
the ABPGM, since bifurcation is impossible, the free end stops and
becomes a \textit{frustrated} dead end. This frustration seems to
be preferable and more realistic than forcing a trifunctional
monomer to turn into bifunctional monomer as it occurs in the
BPGM. So, in our proposal, the relative incidence of bifunctional
and trifunctional monomers does not depend on the topology of the
cluster anymore and is only controlled by the branching
probability parameter. The second difference is also important: in
the BPGM all tips are \emph{sequentially} visited in a clockwise
manner following the sequence of births whereas in the ABPGM they
are visited in a \textit{random} way. The sequential update of the
growth front of the polymer (in the BPGM) simulates the formation
of parallel chains in the infinite phase (for high $b$ \ and low
$c)$ like in a crystallization process. This mechanism leads to a
faceted-to-rough transition that has been already studied
\cite{neves95,lucena99}. Figure 1a shows a typical BPGM
configuration with faceted boundary generated with parameters
$b=0.5$ and $c=0$ (for $L=51$); the parallel ordering of chains is
caused by the clockwise update of the growing tips. If we simulate
the BPGM using the same set of parameters but with a random update
instead of the clockwise update we get Figure 1b. In this case,
two subsequent growing tips are probably located far apart so that
they cannot produce parallel chains. The boundary is less faceted
than before. Recently it
has been shown that any deterministic growth order is \textit{%
non-ergodic} in the sense that it spans only a subset of the space
of all possible configurations \cite{onody00}. Moreover, for the
present purposes, this clockwise update mechanism is undesirable
since the development of parallel chains corresponds to an
increasing incidence of bifunctional monomers due mainly to
geometrical effects rather than to the probability $1-b$  itself.
The relative incidence of bifurcations (which will be defined as
\textit{the effective branching rate }$b_{ef}$ in section 4) is
much smaller than input parameter $b$. Indeed, we anticipate that
$b_{ef}\approx 0.15$ \ for the cluster of Figure 1a whereas
$b_{ef}\approx 0.30$ for the one of Figure 1b; so the last rate
(corresponding to a BPGM cluster generated with random update) is
closer to the input value $b=0.50$. \ The effective branching rate
gets still closer to the input value if we simulate the ABPGM.
Figure 1c is a typical ABPGM configuration (generated with
those same parameters $b=0.5$ and $c=0$)  and presents $b_{ef}\approx 0.43$%
; the faceted front of growth seems to disappear and vacancies can
be found inside the cluster.

\begin{figure}[htbp!]
\begin{center}
\includegraphics[width=5cm]{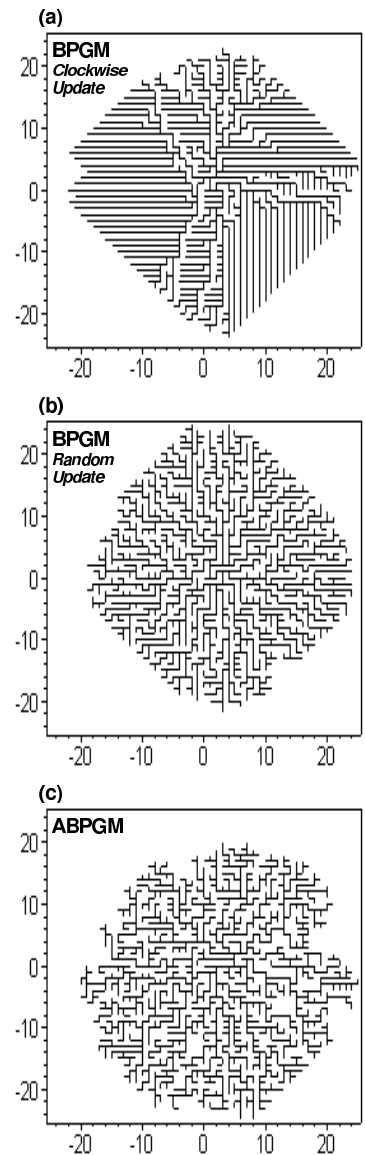}
\end{center}
\caption{Typical configurations simulated on a square lattice of
size $L=51$ with parameters $b=0.5$ and $c=0$ according to the
BPGM rules with clockwise update of growing tips \textbf{(a)},
BPGM with random update \textbf{(b)} and ABPGM \textbf{(c)}. The
corresponding effective branching rates are $b_{ef}\approx 0.15$,
$b_{ef}\approx 0.30$ and $b_{ef}\approx
0.43$, respectively.}%
\label{Figure1}%
\end{figure}

The main feature of the BPGM is the phase transition separating
finite from infinite growth regime. Lucena \textit{et al.}
\cite{lucena94} have defined a critical branching probability
$b_{c}$ where the mean size of finite polymers diverges as
$L\rightarrow \infty $. The critical value depends on the impurity
concentration $c$ and separates the finite phase $(b<b_{c}(c))$
from the infinite one $(b>b_{c}(c))$. The probability that a
polymer grows indefinitely is null for small values of $b$ and
increases abruptly in the region of $b\approx b_{c}(c)$. This
probability was estimated through the \textit{fraction }$P_{\infty
}$\textit{\ of infinite polymers} in the ensemble of
configurations of the BPGM \cite{lucena94}. In Figure 2a, we
reproduce typical plots of $P_{\infty }$ versus $b$ at different
values of $c $ (with $N_{e}=10^{2}$ experiments and $L=1501$) for
the BPGM. In Figure 2b, we present plots of $P_{\infty }$ versus
$b$ at some values of $c$
corresponding to simulations of the ABPGM with $N_{e}=10^{4}$ experiments and $%
L=1501$. For $c=0$, the behavior of $P_{\infty }(b)$ is analogous
to that of the BPGM but now the threshold is little bit higher
($b\approx 0.06$ for the ABPGM while $b\approx 0.055$ for the
BPGM). This difference increases with $c$. The most interesting
characteristic of the present model may be observed when $c=0.18$.
For this value, the curve raises at $b\approx 0.32$, then presents
a plateau (where $P_{\infty }(b)\approx 0.7$) and finally falls!
By a finite-size scaling analysis we have verified that the height
of the plateau does not change significatively in the limit
$L\rightarrow \infty .$ For $c=0.19$, the behavior of $P_{\infty
}(b)$ is gaussian shaped. This means that, for certain impurity
concentrations, as $b$ increases the system goes from a finite to
an infinite phase and then becomes finite again! Indeed, this
reentrance is confirmed in the next section when we determine the
phase diagram of the ABPGM through an analysis of the correlation
length.

\begin{figure}[htbp!]
\begin{center}
\includegraphics[width=6.5cm]{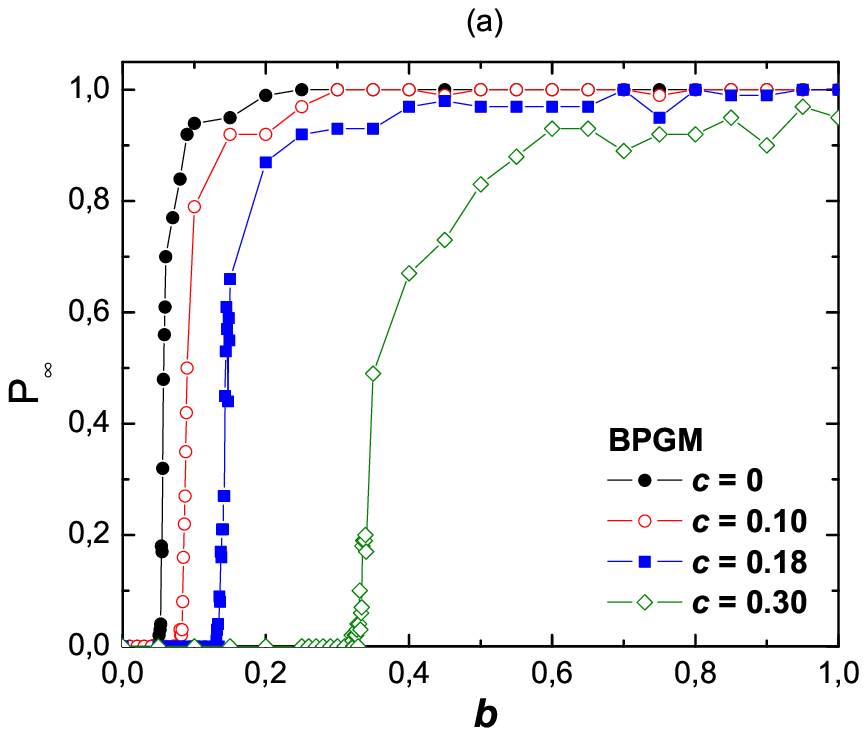}
\includegraphics[width=6.5cm]{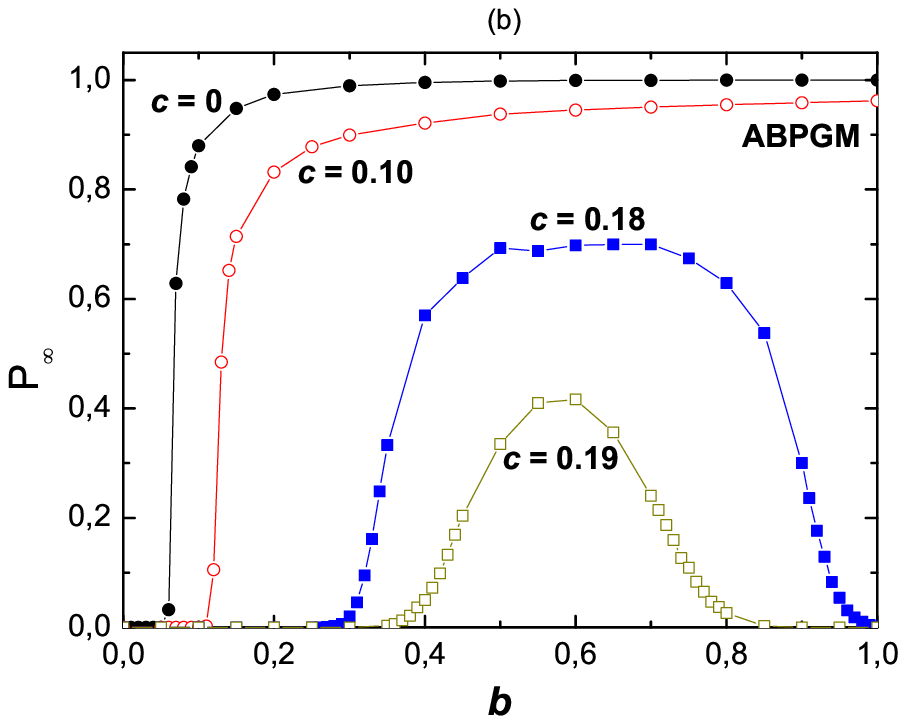}
\end{center}
\caption{Plots of the fraction of infinite polymers $P_{\infty}$
versus $b$ for the BPGM \textbf{(a)} and ABPGM
\textbf{(b)}.}%
\label{Figure2}%
\end{figure}

\section{The Phase Diagram}

The mean size of finite polymers is a measure of the \textit{%
correlation length }$\xi $ of the system. If a finite polymer is
generated during a simulation, the sizes $l_{x}$ and $l_{y}$ of
the smallest rectangle containing the cluster can be determined.
The correlation length can then be calculated as $\xi =\langle
\left( l_{x}l_{y}\right) ^{1/2}\rangle $ where the average is
performed over all experiments with finite polymers
\cite{lucena94}. We show typical plots of $\xi $ versus $b$
corresponding to simulations of the BPGM (in Figure 3a) and ABPGM
(in Figure 3b) for different values of $c$ and size $L=1501$ (with
$10^{2}$ and $10^{4}$ experiments respectively). Each plot of $\xi
$ versus $b$ exhibits only one maximum except for plots of the
ABPGM with $c$ $>0.175$ where two peaks are
detected! It can be verified that all peaks of $\xi $ do diverge when $%
L\rightarrow \infty $.

\begin{figure}[htbp!]
\begin{center}
\includegraphics[width=6cm]{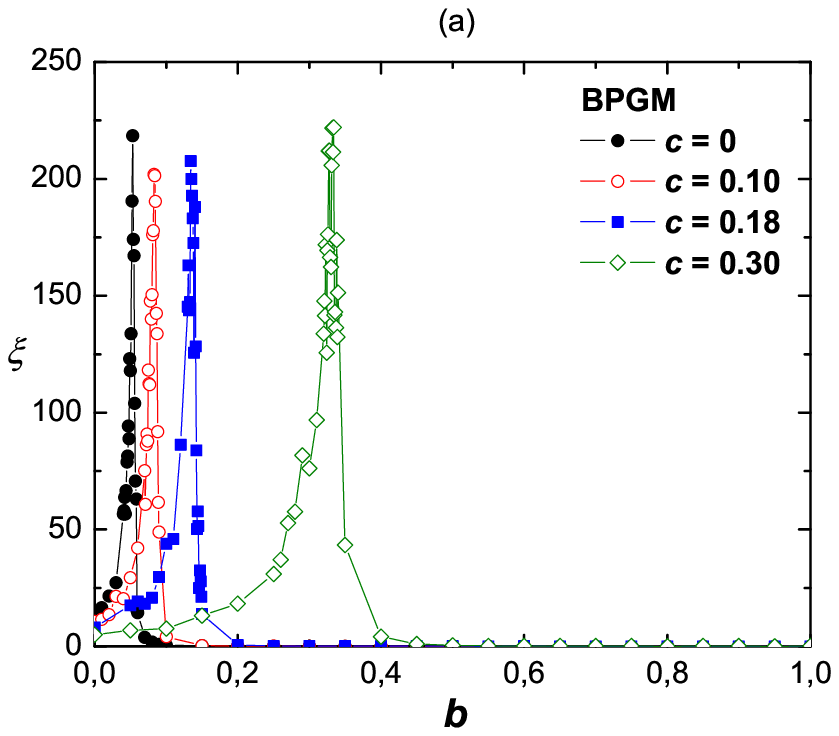}
\includegraphics[width=6cm]{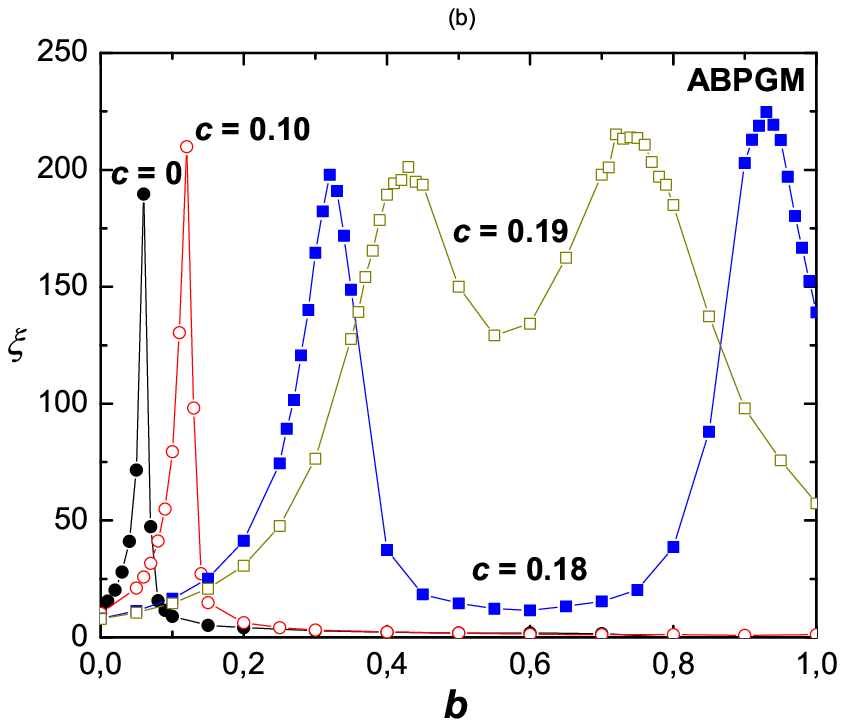}
\end{center}
\caption{Plots of the correlation length $\xi $ of finite polymers
versus the branching probability $b$ for several impurity
concentrations $c$ in the cases: \textbf{(a)} BPGM and
\textbf{(b)} ABPGM.}%
\label{Figure3}%
\end{figure}

Let us first explain the behavior of $\xi $ for the BPGM. For a fixed value of $%
c $ and as $b$ increases, the polymer is more likely to escape
from steric hindrances and impurities so that the mean size grows
to its highest value at some $b=b_{0}(L)$. Above this point, the
system is defined to be in the infinite growth regime (of course,
the true critical point is obtained in the thermodynamic limit
$b_{c}=\lim_{L\rightarrow \infty }b_{0}(L)$). As $b$ continues to
increase, the fraction of infinite polymers grows and the finite
polymers get smaller so that $\xi $ decreases. As higher impurity
concentrations hindrance the growth, $b_{0}$ increases with $c$.
The critical line of the BPGM is the locus $(b_{c},c)$ on which
$\xi $ diverges \cite{lucena94} and is shown in Figure 4 (dashed
line).

Regarding the ABPGM, the same reasoning can explain the maximum of
$\xi $ (or the first maximum when there are two peaks). But now
this peak is located on a higher branching probability
$b_{1}>b_{0}$ that compensates the occurrence of frustrated dead
ends. Just above $b_{1}$ the system enters the infinite
growth regime\ where it remains unless a second peak appears (at $%
b=b_{2}>b_{1}$ for $c$ $>0.175$). In the latter case the system
returns to the finite phase! This reentrance from infinite to
finite growth regime is a peculiar feature of the ABPGM. For this
modified model, there is a certain range of values of $c$ where it
is very probable that all free ends become frustrated (and stop
growing) for a sufficiently large $b$ so that the polymer growth
is finite again. The ABPGM phase diagram is also shown in the
Figure 4. The reentrant phase only exists for $c$ in the small
interval $[0.175,0.195]$.

\begin{figure}[htbp!]
\begin{center}
\includegraphics[width=6cm]{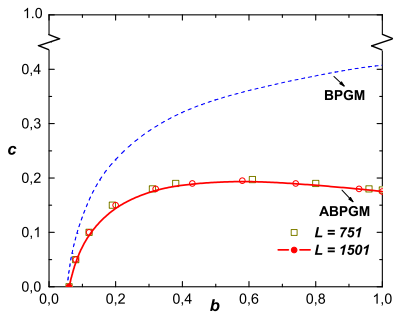}
\end{center}
\caption{The phase diagram of the models BPGM (dashed line) and
ABPGM (solid line). Each critical line separates the infinite
phase (at low values of $c$) from the finite one (at
higher values of $c$).}%
\label{Figure4}%
\end{figure}

\begin{figure}[htbp!]
\begin{center}
\includegraphics[width=6cm]{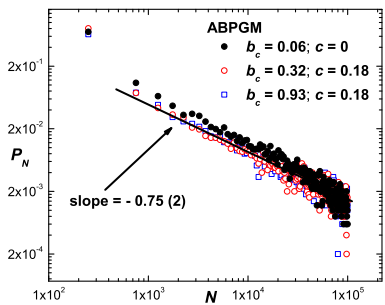}
\end{center}
\caption{Log-log plot of the distribution $P(N)$ of finite
polymers with $N$ bonds for simulations of the ABPGM (with
$N_{e}=10^{4}$ and $L=1501$) at some points $(b,c)$ on the
critical line.}%
\label{Figure5}%
\end{figure}

We have also measured the fraction $P(N)$ of finite polymers with
$N$ bonds. In Figure 5, we have a log-log plot of the
polydispersion distribution $P(N)$ of the ABPGM on the critical
line
(running $10^{4}$ experiments in a $%
L=1501$ lattice). The three sets of data correspond\ to the critical points: $%
b_{1}\approx 0.06$ and $c=0$ (black dots); $b_{1}\approx 0.32$ and
$c=0.18$ (open circles); $b_{2}\approx 0.93$ and $c=0.18$
(squares). The data are fitted by a straight line with slope
$-0.75$. So, on the critical line, $P(N)$ decays with $N$ as a
power law. We have verified that outside the critical line, $P(N)$
decays exponentially with $N$. Lucena \textit{et al.}
\cite{lucena94} have found a similar behavior in the BPGM.

\section{The Frustration and the Effective Branching Rate}

The \textit{frustration} is a desirable event which distinguishes
the ABPGM from the original model. It is defined as the
\textit{interruption} of the growth of any free end which was
chosen (with probability $b$) as a trifunctional monomer but is
unable to bifurcate since only one nearest neighbor site is
available. It prevents such a free end to continue its growth
linearly like a bifunctional monomer (as occurs in the BPGM) and
consequently controls the relative incidence of branches.

\begin{figure}[htbp!]
\begin{center}
\includegraphics[width=6cm]{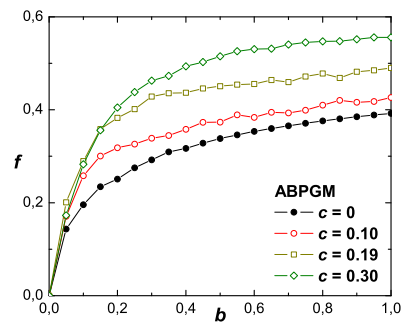}
\end{center}
\caption{The \textit{frustration rate} $f$ versus $b$ for
increasing values of $c$ in the ABPGM ($L=1501$ and $N_{e}\sim
10^{3}$).}%
\label{Figure6}%
\end{figure}

Clearly, each polymer configuration of the ABPGM is also a bond
tree which
can be classified by the numbers $N$ (of bonds) and $N_{k}$ (of $k-$%
functional units). Besides, every \textit{dead end} is a site
which ceased to grow and so represents a monofunctional vertex; it
can be subclassified
as either a \textit{trapped site} (if it is in a "cul de sac") or a \textit{%
frustrated dead end }(otherwise). We denote by $\widetilde{N}_{1}$
the number of frustrated dead ends. For the BPGM,
$\widetilde{N}_{1}=0$ since any free end never stops. In order to
measure the relative incidence of frustrated ends among those free
sites \textit{chosen} as trifunctional monomers, we define the
\textit{frustration rate} as

\begin{eqnarray}
f=\left\langle \frac{\widetilde{N}_{1}}{\widetilde{N}_{1}+N_{3}}%
\right\rangle
\end{eqnarray}
where the average is performed over \textit{all} experiments. For
infinite polymers, the current sites on the front of growth should
not be considered in the computation since their functionalities
are undetermined.

According to the prior definition, $f=0$ for any values of $b$ and
$c$ in the BPGM. We remark that the null frustration of the BPGM
does \textit{not} mean that all free sites \textit{chosen} to
bifurcate succeed but only that when they fail they are
transformed in monomers with functionality two.

For the ABPGM, the frustration rate increases with $b$ and $c$ as
it is shown in Figure 6. Indeed, both the excluded volume due to
self-avoidance (which increases with $b$) and the impurities
diminish the chance of success of any free end, so $f$ is a
monotonically increasing function of $b$ and $c$.

\begin{figure}[htbp!]
\begin{center}
\includegraphics[width=6.5cm]{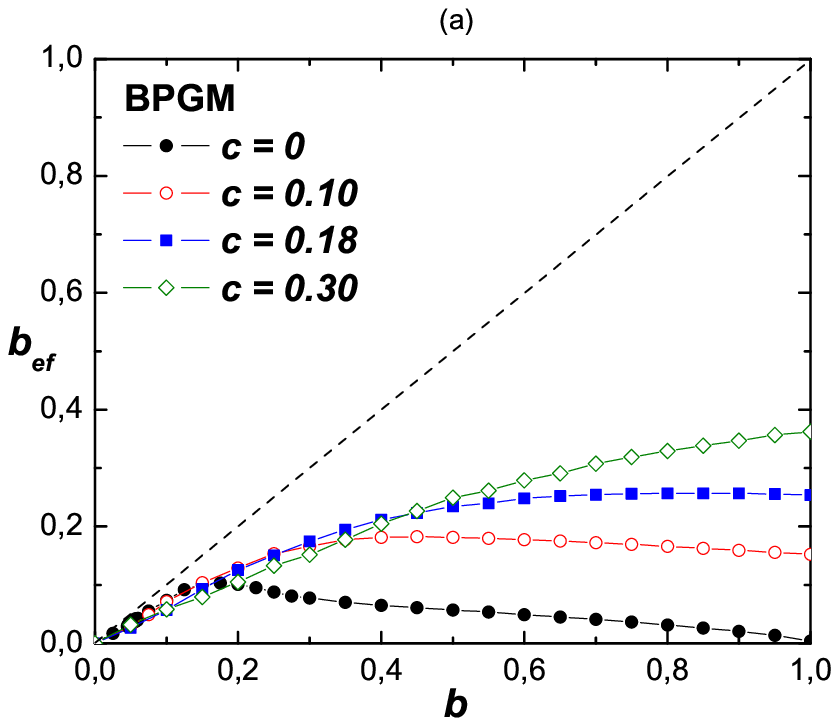}
\includegraphics[width=6.5cm]{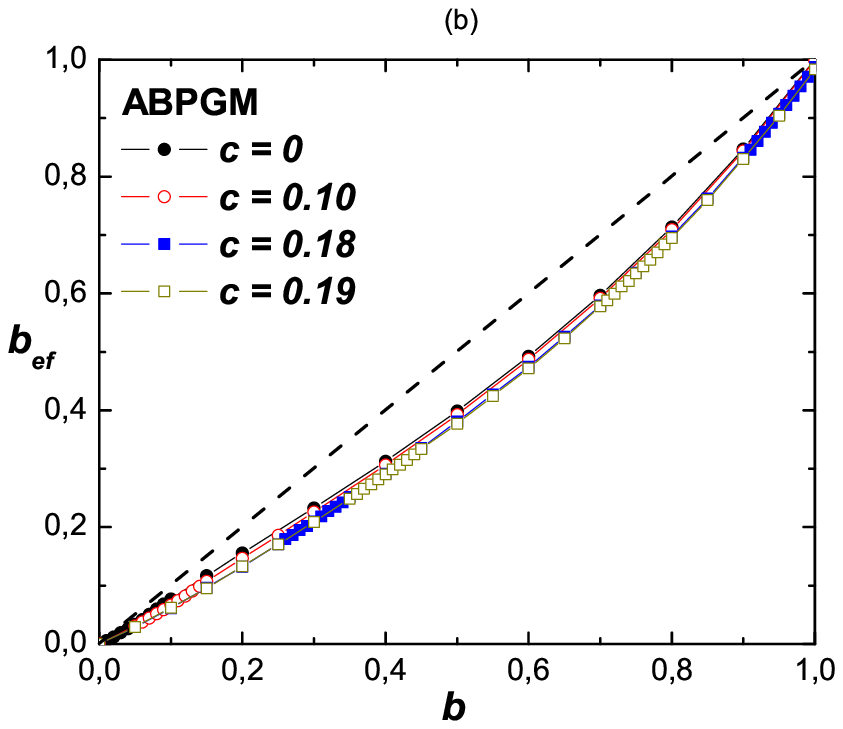}
\end{center}
\caption{The behavior of the \textit{effective branching rate}
$b_{ef}$ versus the input parameter $b$ for simulations of the
BPGM \textbf{(a)} and ABPGM \textbf{(b)} on a large $L=1501$
square lattice (with $10^{2}$ and  $10^{4}$ experiments
respectively). The dashed lines represent the ideal behavior
$b_{ef}=b$.}%
\label{Figure7}%
\end{figure}

The effectiveness of the input parameter $b$ is evaluated by
comparing it with the relative frequency of branches in a polymer
configuration. For this purpose, we define the \textit{effective
branching rate} $b_{ef}$ as the ratio between $N_{3}$ and
$N_{2}+N_{3}$ averaged over all experiments:

\begin{eqnarray}
b_{ef}=\left\langle \frac{N_{3}}{N_{2}+N_{3}}\right\rangle
\end{eqnarray}

The plots of $b_{ef}$ versus $b$ at different values of $c$ for
simulations of the BPGM and ABPGM (on a large $L=1501$ square
lattice) are shown in Figures 7a and 7b, respectively. The ideal
behavior $b_{ef}=b$ is indicated as the dashed lines. Regarding
the BPGM, there is evidently a large discrepancy between the input
parameter $b$ and the output rate $b_{ef}$. For $c=0$, $b_{ef}$
increases with $b$ up to a maximum (at $b\approx 0.2$) and then
decreases to
zero as $b\rightarrow 1$. This behavior is explained as follows: for small $%
b$, self-avoidance is reduced so that most sites trying to
bifurcate succeed; but as $b$ increases, parallel linear chains
(like those of Figure 1a) are forcibly generated due to both the
increasing excluded volume and the clockwise update. For higher
values of $c$, the presence of impurities hampers the formation of
linear chains and thus $b_{ef}$ decreases slower (as a
\textit{concave} function). Anyway, for
the BPGM, the discrepancy between $b$ and $b_{ef}$ gets more pronounced as $%
b\rightarrow 1$. On the contrary, for the ABPGM, $b_{ef}$ always increases with $%
b$ as a \textit{convex} function (for any value of $c$) and the difference\ $%
b-b_{ef}$ is very small. In fact, for any $c$, the ratio
$b_{ef}/b$ is approximately equal to $1$ for small $b$, decreases
until about $0.8$ for intermediate $b$ and then returns to $1$ as
$b\rightarrow 1$. Such results corroborate the assertion that the
input parameter $b$ controls the relative incidence of branches in
the adapted model.

\begin{figure}[htbp!]
\begin{center}
\includegraphics[width=5.0cm]{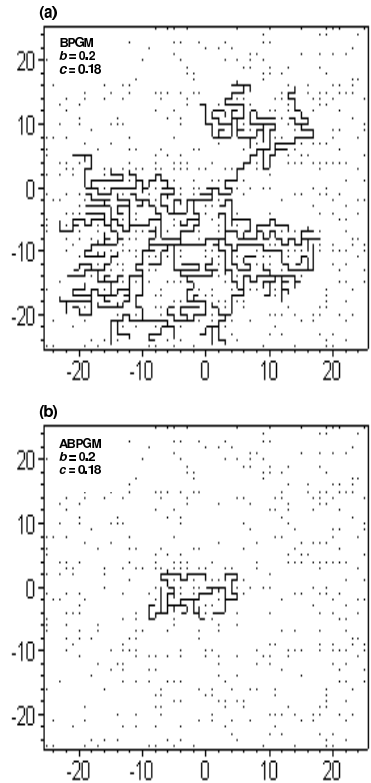}
\end{center}
\caption{Typical graphs of both the BPGM \textbf{(a)} and
ABPGM \textbf{(b)} for $b=0.2$ and fixed $c=0.18$.}%
\label{Figure8}%
\end{figure}

\begin{figure}[htbp!]
\begin{center}
\includegraphics[width=5.0cm]{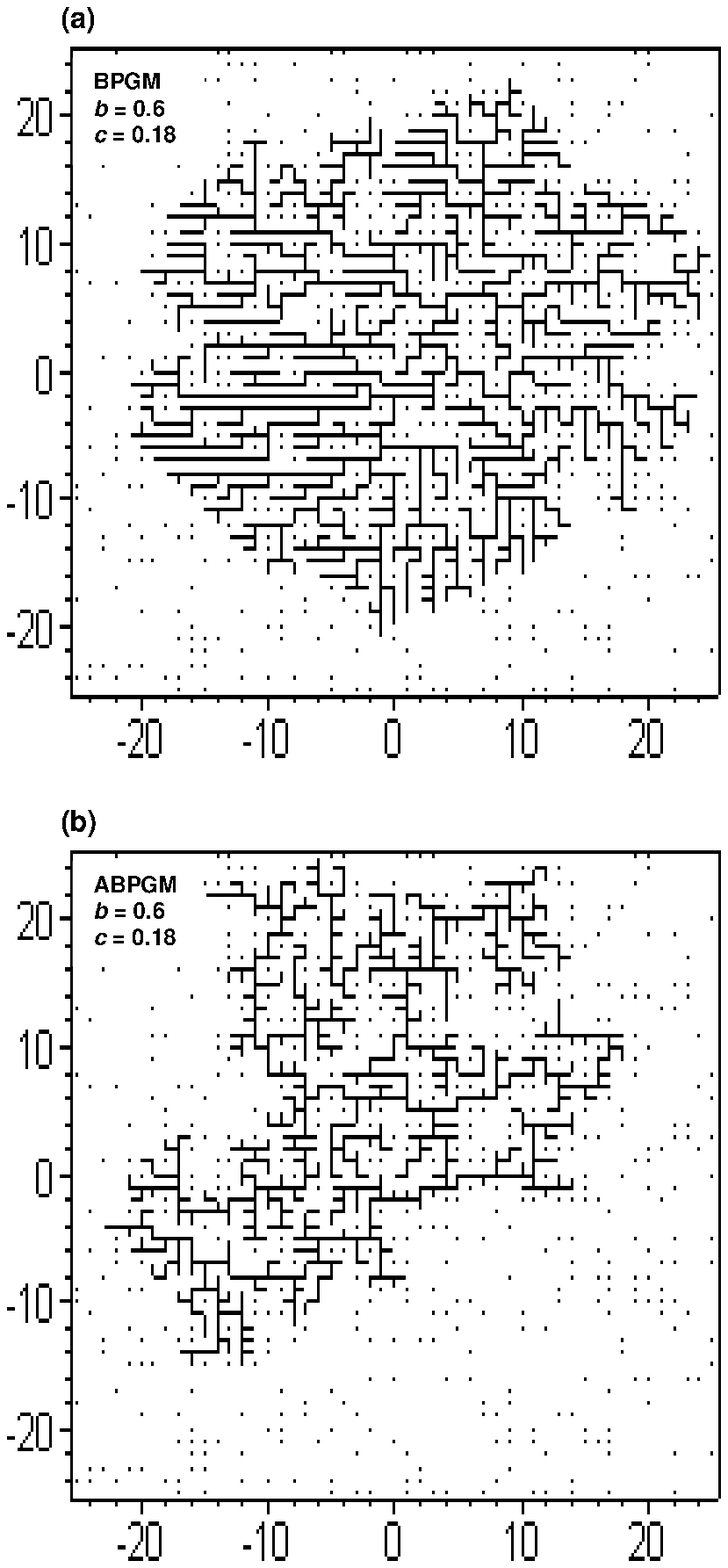}
\end{center}
\caption{Typical graphs of both the BPGM \textbf{(a)} and
ABPGM \textbf{(b)} for $b=0.2$ and fixed $c=0.18$.}%
\label{Figure9}%
\end{figure}

\begin{figure}[htbp!]
\begin{center}
\includegraphics[width=5.0cm]{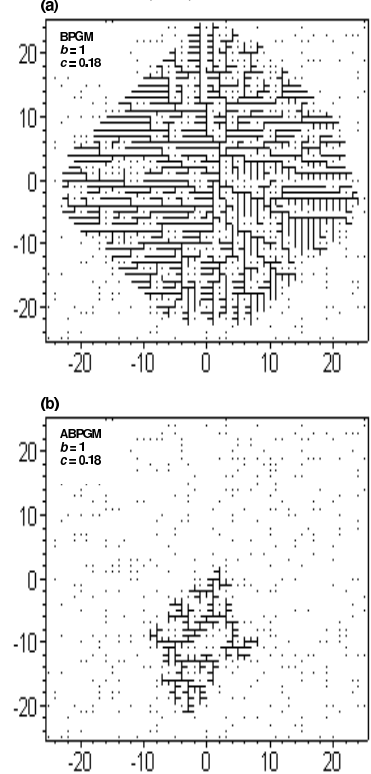}
\end{center}
\caption{Typical graphs of both the BPGM \textbf{(a)} and
ABPGM \textbf{(b)} for $b=1$ and fixed $c=0.18$.}%
\label{Figure10}%
\end{figure}

\section{Conclusions}

The \textit{branched polymer growth model} (BPGM) was originally
proposed \cite{lucena94} as a generalization of the
\textit{kinetic growth walk} \cite{majid84} in order to include
the possibility of ramification of the polymer as well as the
presence of impurities in the medium. The model was found to
exhibit a finite-infinite transition due to competition between
branching and hindrances.

In this paper, we have proposed an alteration in the dynamics of
the BPGM so as to \textit{adapt} the model to an experimental
realism. We have called it the \textit{adapted branched polymer
growth model} (ABPGM). The main difference between our proposal
and the original model regards to the growth mechanism of a
monomer which is chosen to bifurcate (with probability $b$) but
has just one empty nearest neighbor site. In the BPGM, such a
monomer is transformed into a bifunctional unit so that it grows
linearly (with probability one). In our adapted model, that
monomer stops and becomes a frustrated dead end. This frustration
reveals as preferable and more realistic than changing the monomer
functionality from $3$ to $2$ (as it occurs in the BPGM). This
subtle change in the algorithm together with a \textit{random}
update of the growing ends lead to the formation of polymers with
new topological patterns and adjusted degrees of branching.
Indeed, we have shown that the effective branching rate $b_{ef}$
is very much closer from the input parameter $b$ in the ABPGM than
in the original model.

We have found that the ABPGM presents a finite-infinite transition
in the $(b,c)$ space with a peculiar reentrant phase in the small
interval $0.175\leq c\leq 0.195$. At this instance, we compare
some typical graphs of both the BPGM and ABPGM at the
\textit{fixed} impurity concentration $c=0.18$ for three
increasing values of $b$: $0.2$, $0.6$ and $1$ (Figures 8, 9 and
10, respectively). For the BPGM, all the three corresponding
configurations are \textit{infinite} clusters whose boundaries
change from rough to faceted as $b$ increases. On the other hand,
the typical ABPGM graph for $b=0.2$ is a \textit{finite} cluster
(due to the occurrence of \textit{both} trapped sites and
frustrated dead ends); if $b=0.6$ the cluster is \textit{infinite}
(since here higher branching overcomes dead ends) and if $b=1$ the
cluster is \textit{finite} again (due to a high frustration rate).

We hope that the ideas presented here can add a new degree of
physical understanding to the model as well as they can
approximate it to the real ramified polymers.

This work has been supported by Brazilian Agencies CAPES, CNPq and
FAPESP (through grant No. 96/05387-3).

\end{document}